\documentclass[aps,pra,twocolumn,nopacs,showemail]{revtex4}
\usepackage{bm}
\usepackage{amsmath}
\usepackage{latexsym}
\usepackage{color}
\usepackage{tensor}
\usepackage{braket}
\usepackage{graphicx}
\usepackage{siunitx}
\usepackage{slashed}
\usepackage{dcolumn}   
\newcommand{\abs}[1]{\left\lvert#1\right\rvert}
\newcommand{\as}[1]{\renewcommand{\arraystretch}{#1}}
\newcommand*{\centt}[1]{\multicolumn{1}{c}{#1}}
\newcommand*{\cent}[1]{\multicolumn{1}{c}{$#1$}}
\newcolumntype{w}[1]{D{.}{.}{#1}}

\newcommand{\nn}{\nonumber}
\allowdisplaybreaks

\begin{document}
\preprint{Version 3.0}

\title{Nuclear-structure corrections to the hyperfine splitting in muonic deuterium}

\author{Marcin Kalinowski}
\email[]{mj.kalinowski@student.uw.edu.pl}
\affiliation{Faculty of Physics,
	University of Warsaw,
	Pasteura 5, 02-093 Warsaw, Poland}
\author{Krzysztof Pachucki}
        \homepage[]{www.fuw.edu.pl/~krp} \affiliation{Faculty of Physics,
	University of Warsaw,
	Pasteura 5, 02-093 Warsaw, Poland}
\author{Vladimir ~A. Yerokhin}
\affiliation{Center for Advanced Studies, Peter the Great St.~Petersburg Polytechnic University,
195251 St.~Petersburg, Russia}

\date{\today}

\begin{abstract}
  Nuclear structure corrections of orders $Z\alpha\, E_F$ and $(Z\alpha)^2 E_F$ are calculated
  for the hyperfine splitting of the muonic deuterium. The obtained results disagree with
  previous calculations and lead to a $5\,\sigma$ disagreement with the current experimental
  value of the $2S$ hyperfine splitting in $\mu$D.
\end{abstract}

 \maketitle

\section{Introduction}

Nuclear structure effects represent the main limitation for precise theoretical description of
atomic energy levels. These effects are particularly important for muonic atoms, where the Compton
wavelength of the bound muon $\sim\!\!2$~fm is of the same order as the nuclear size. It is,
therefore, not surprising that the uncertainty of modern theoretical predictions of energy levels
of light muonic atoms is dominated by the uncertainty of nuclear-structure effects. Specifically,
the current theoretical value of the hyperfine splitting (hfs) of the $2S$ state of muonuic deuterium ($\mu$D) is
\cite{krauth:16}
\begin{equation}\label{eq1}
  E_{\rm hfs}(2S)_{\rm theo} = 6.2791\,(50)\,{\rm meV}\,,
  \end{equation}
where the uncertainty comes almost exclusively from the deuteron polarizability ($\pm 0.0049$~meV).
The theoretical value (\ref{eq1}) was obtained in Ref.~\cite{krauth:16} by compiling two
independent calculations, by Borie~\cite{borie:14} and by Faustov {\em et
al.}~\cite{faustov_hyperfine_2014}. The theoretical result is in good agreement with the
experimental value \cite{pohl:16}:
\begin{equation}\label{eq:Enuclexp}
  E_{\rm hfs}(2S)_{\rm exp} = 6.2747(70)_{\rm stat} (20)_{\rm syst}\;{\rm meV}\,.
  \end{equation}

In this work we will demonstrate that the deuteron structure corrections to hfs in $\mu$D were
previously treated incorrectly and that the good agreement with the experimental value was probably
accidental. Specifically, the calculation by Borie \cite{borie:14} included only the elastic part
of the nuclear structure, which does not reproduce even the correct sign of the total nuclear
structure effect. Faustov and coworkers \cite{faustov_hyperfine_2014} included both the Zemach and
the Low corrections at the same time, which is inconsistent, and, moreover, used an incorrect
formula for the nuclear polarizability.

In the present work we derive the nuclear-structure corrections in $\mu$D induced by the two- and
three-photon exchange between the bound muon and the nucleus and analyze discrepancies with the
previous calculation \cite{faustov_hyperfine_2014} and the experimental result \cite{pohl:16}.

The leading-order hfs of atomic levels of light atoms is independent of nuclear structure and is
given (for the $nS$ states) by the Fermi contact term,
\begin{equation}\label{eq:Ef}
E_F = \frac{4g_N m_r^3}{3 m_p\,m\,n^3}(Z\alpha)^4\braket{\vec{s}_N\cdot\vec{s}_\mu}\,,
\end{equation}
where $n$ is the principal quantum number, $Z$ is the nuclear charge number, $\alpha$ is the
fine-structure constant, $m$ and $m_p$ are masses of the lepton and the proton, respectively,
$m_r$ is the reduced mass of the atom, $\vec{s}_\mu$ and $\vec{s}_N$ are the spin operators of the
lepton and the nucleus, respectively, $ g_N $ is the 'modified' $g$-factor of the nucleus defined by
\begin{equation}
  \vec \mu_N = \frac{Z e\,g_N}{2\,m_p}\,\vec{s}_N\,,
\end{equation}
$\vec \mu_N$ is the nuclear magnetic moment operator and $e$ is the elementary charge.
Numerical values of $E_F$ for the ground and the first excited state of muonic deuterium are
\begin{align}
  E_F = \left\{
  \begin{array}{ll}
    \SI{49.0875}{\milli\electronvolt}\,, & \mbox{\rm for the $1S$ state}\,,\\
    \SI{6.13594}{\milli\electronvolt}\,, & \mbox{\rm for the $2S$ state}\,.
  \end{array}\right.
\end{align}
If we assume the point nuclear model and account for all known QED corrections, the theoretical
result for the $2S$ state is \cite{krauth:16}
\begin{equation}\label{eq6}
  E_{\rm hfs}({\rm point}) = \SI{6.17815\pm0.00020}{\milli\electronvolt}\,,
\end{equation}
which corresponds to the sum of entries
$h_1+h_2+h_4+h_5+h_7+h_8+h_9+h_{9b}+h_{12}+h_{13}+h_{14}+h_{18}$ from Table IV of
Ref.~\cite{krauth:16}. The deviation of the experimental value (\ref{eq:Enuclexp}) from the
theoretical point-nucleus result (\ref{eq6}) can be regarded as
the ``experimental value'' of the nuclear-structure correction,
\begin{equation} \label{eq:Ediff}
  \delta E_{\rm nucl} =  E_{\rm hfs}({\rm exp}) - E_{\rm hfs}({\rm point}) = \SI{0.0966\pm0.0073}{\milli\electronvolt}\,.
\end{equation}

From the theoretical side the nuclear-structure correction for light atoms
can be described within the $Z\,\alpha$ expansion,
\begin{equation}\label{eq:Enucl}
  \delta E_{\rm nucl} =   \delta^{(1)} E_{\rm nucl} +   \delta^{(2)} E_{\rm nucl} + \ldots
\end{equation}
where $\delta^{(1)} E_{\rm nucl}$ is the two-photon exchange correction of order
$(Z\,\alpha)\,E_F$, $\delta^{(2)} E_{\rm nucl}$ is the three-photon exchange correction of order
$(Z\,\alpha)^2\,E_F$, and $\ldots$ denotes smaller contributions due to exchange of
larger number of photons and radiative corrections.
In the following discussion, we calculate the two-photon and the
three-photon exchange nuclear-structure corrections for the hfs of $nS$ states of muonic deuterium.
Relativistic units ($\hbar = c = 1$) are employed throughout.

\section{Two-photon exchange nuclear structure }
The most straightforward way of including the nuclear effects is to assume that the nucleus is
described by some elastic electric and magnetic formfactors. This leads to the so-called elastic, or
finite-nuclear-size (fns) corrections. The leading fns correction originates from the two-photon
exchange. It was derived long ago by Zemach \cite{zemach_proton_1956} and is given by
\begin{equation}\label{eq:Ezem}
\delta E_{\text{Zem}} = -2\, m_r Z\alpha\, r_Z E_F,
\end{equation}
where $ r_Z $ is the Zemach radius defined by
\begin{equation}\label{eq:rZ}
r_Z = \int d^3r_1 \int d^3r_2 \,\rho_M(r_1)\,\rho_E(r_2)\,\abs{\vec{r}_1-\vec{r}_2}\,.
\end{equation}
(Note that the subscript $ Z $ in $r_Z$ is not related to the nuclear charge.) In the above
equation, $ \rho_E $ and $ \rho_M $ are the charge and the magnetic-moment distributions of the
nucleus, respectively, i.e., the Fourier transform of the corresponding formfactors. The numerical
value of the Zemach correction for the $2S$ state of $\mu$D, in the nonrecoil limit, with $r_Z =
2.593(16)$~fm~\cite{friar_sick:04}, is
\begin{equation}
\delta E_{\text{Zem}} = \SI{-0.1177(33)}{\milli\electronvolt}\,.
\end{equation}
We note the opposite sign of the Zemach correction as compared to the experimental value of the
total nuclear structure \eqref{eq:Ediff}. This demonstrates that the description of the nucleus
only through the elastic formfactors is not adequate.

Various hfs corrections arise from excitations of the nucleus by the bound lepton, usually referred
to as the {\em inelastic} nuclear-structure corrections. In the case of the proton, the inelastic
contribution can be obtained from the experimentally accessible spin-dependent structure functions
by using dispersion relations
\cite{tomalak_two-photon_2017,carlson_proton_2008,carlson_proton_2011,tomalak_hyperfine_2018}. For
other nuclei, including deuteron, the inelastic spin-dependent structure functions are unknown, and
one has to rely on theoretical calculations.

In our calculations we consider the elastic and inelastic contributions together, and use
a perturbation expansion over a small parameter, namely, the
ratio of the average nucleon binding energy over the nucleon mass. Specifically, the two-photon
exchange correction can be represented as
\begin{equation}\label{eq:E2pe}
\delta^{(1)} E_{\rm nucl} = \delta E_{\rm Low} + \delta E_{\rm 1nucl} + \delta E_{\rm pol} +\ldots\,.
\end{equation}
The leading-order  term $\delta E_{\rm Low}$ was first derived by Low \cite{low_effects_1950}. For
the particular case of an $ nS $ state in $\mu$D, the numerical value of Low's correction in the
point-nucleon model is
\begin{equation} \label{approx}
\delta E_{\rm Low} \approx -2\,m_r\,\alpha\,E_F\,\frac{g_n}{g_d}\langle R\rangle = \frac{2.640}{n^3}\,{\rm meV}\,.
\end{equation}
Here, $R$ is the distance of the proton from the center of mass. Its expectation value was
calculated using the AV18 potential \cite{wiringa:95} as $ \Braket{R} = \SI{1.629}{\femto\meter}$.

A more detailed calculation of the leading-order term was performed by Friar and Payne
\cite{friar_nuclear_2005,friar_nuclear_2005-1}, with inclusion of the finite nucleon size and meson
exchange currents. Their result was reported for the $1S$ state of ordinary (electronic) deuterium,
$\delta E_{\rm Low}(e{\rm D}) = \SI{87.3}{\kilo\hertz}$. Rescaling it to the $nS$ states of $\mu$D one gets
\begin{equation}\label{eq:Elow}
  \delta E_{\rm Low} = \frac{m_r^4(\mu D)}{m_r^4(e D)}\frac{m(e)}{m(\mu)}\frac{1}{n^3}\,\delta E_{\rm Low}(eD) = \frac{2.566}{n^3}\,{\rm meV}\,,
\end{equation}
in good agreement with the approximate result of Eq.~(\ref{approx}).

$\delta E_{\rm 1nucl}$ is the contribution induced by individual nucleons.
It is given by the individual nucleon Zemach corrections,
\begin{equation}\label{eq15}
\delta E_{\rm 1nucl} = -\frac{8\,\alpha^2}{3\,n^3}\,\frac{m_r^3}{m_p+m}\,
\vec s_\mu\cdot\bigl\langle\sum_a g_a\,\vec s_a\,r_{a\rm Z}\bigr\rangle.
\end{equation}
where $r_{aZ}$ is the effective radius of the nucleon $a$. If only the elastic part is included,
the result for the proton is $r_{pZ} = \SI{1.086 (12)}{\femto\meter}$ \cite{friar_sick:04} and for the neutron,
$r_{nZ} = -0.042$ fm \cite{friar_nuclear_2005-1}.
The result for the proton effective radius that includes the recoil and polarizability
contributions can be obtained from the $1S$ muonic hydrogen correction
$\delta E_{\rm nucl}(\mu H) = \SI{-1.131(24)}{\milli\electronvolt}$
obtained by Tomalak in Ref. \cite{toamalak_arxiv_18}, namely $r_{p\rm Z} = 0.883(19)$\,fm,
and for the neutron from Ref. \cite{tomalak}, $r_{n\rm Z} = 0.06(1)$ fm.

$\delta E_{\rm pol}$ is the contribution of the nuclear vector polarizability. It has been studied
for ordinary atoms by Khriplovich and Milstein \cite{khriplovich_corrections_2004} and later by one
of us (Pachucki) \cite{pachucki_nuclear_2007}. Results of Ref.~\cite{khriplovich_corrections_2004}
obtained in the logarithmic approximation were shown in Ref.~\cite{pachucki_nuclear_2007} to be
incorrect, because the coefficient of logarithm for an arbitrary chosen cutoff vanishes completely.
Moreover, the derivation of Ref.~\cite{khriplovich_corrections_2004} is applicable only for the
'electronic' atoms but not for the muonic ones. Nevertheless, the result of
Ref.~\cite{khriplovich_corrections_2004} was later used by Faustov {\em et al.} in their
calculations of the nuclear structure in $\mu$D \cite{faustov_hyperfine_2014}. In the present work
we derive the nuclear vector polarizability correction $\delta E_{\rm pol}$ for muonic deuterium.

Following the method of Ref.~\cite{pachucki_nuclear_2007},
we obtain the vector polarizability corrections to hfs in $\mu$D in the form
\begin{align}
\delta E_{\rm pol} =&\  -e^2\,\psi^2(0)\,\int\frac{d\,\omega}{2\,\pi}\,
\int\frac{d^3k}{(2\,\pi)^3}\label{22} \\ &\hspace*{-5ex}
\times
\frac{\bigl(\omega^2\,\epsilon^{klj}+k^i\,k^k\,\epsilon^{lij}
-k^i\,k^l\,\epsilon^{kij}\bigr)
\,\sigma^j\,\alpha^{kl}}{\omega\,(\omega^2 + 2\,m\,\omega-k^2)\,
  (\omega^2 - 2\,m\,\omega-k^2)\,(\omega^2 - k^2)}\,,\nonumber
\end{align}
where $\alpha^{kl}$ is the antisymmetric part of the scattering amplitude tensor $T^{kl}$. In the
simplest case of the electric dipole coupling $-\vec D\cdot \vec E$, $\alpha^{kl}$ takes the form
\begin{equation}
\alpha^{kl} = \omega^2\,
\biggl\langle D^k\frac{1}{E_N-\tilde H_N-\omega}\,D^l +
D^l\frac{1}{E_N-\tilde H_N+\omega}\,D^k\biggr\rangle,
\end{equation}
where $\tilde H_N = H_N+k^2/(2\,M)$,  $H_N$ is the nuclear internal Hamiltonian,
$E_N = \Braket{H_N}$, $\vec D = \sum_a e_a \vec R_a$,
and $\vec R_a$ is the nucleon position with respect to the mass center. After integration over
$\omega$ and $k$, and expansion in the small parameter
\begin{equation}
X = \sqrt{\frac{2(H_N-E_N)}{m_r}}\,,
\end{equation}
we obtain the dipole polarizability correction as
\begin{equation}\label{eq:Epol0}
\delta E_{\text{pol0}} = -\frac{i\,e^2}{12\pi}\psi^2(0)\,\Bigl(\frac{m_r}{m}\Bigr)^2\,\epsilon^{ijk}\sigma^k\Braket{D^i\,X\,D^j}\,.
\end{equation}
$\delta E_{\text{pol0}}$ vanishes in the nonrelativistic limit and its numerical contribution is
expected to be small, because it requires the presence of both the spin-orbit and the
quadrupole-spin interactions between nucleons.

There are, however, other polarizability corrections that yield significant numerical
contributions. The first one is the correction due to the magnetic quadrupole interaction,
\begin{equation}\label{eq:dHmag}
\delta H = -\sum_a\frac{ e\,g_a}{2\,m_p}\,R_a^i\,s_a^j\,B^j_{,i}\,.
\end{equation}
The corresponding contribution to the scattering tensor is
\begin{align}
&\delta \alpha^{kl} = i\,\omega\, \sum_a\,\frac{e\,g_a}{2\,m_p}\,
\biggl\langle
\vec R_a\vec k\,(\vec s_a\times\vec k)^k\,\frac{1}{E_N-\tilde H_N-\omega}\,D^l
\nonumber \\ &
+D^l\,\frac{1}{E_N-\tilde H_N+\omega}\,\vec R_a\vec k\,(\vec s_a\times\vec k)^k
-(k\leftrightarrow l, \omega\rightarrow -\omega)\biggr\rangle\,.\label{eq:vpEpol1}
\end{align}
The first term in the small-$X$ expansion gives the following correction to the hyperfine splitting
of $\mu$D
\begin{align}\label{eq:Epol1}
\delta E_{\text{pol1}} =& -\frac{2\alpha}{3}\frac{g_p-g_n}{g_d}E_F\,
 m_r^2\Braket{\vec{R}\,X\,\vec{R}}\,.
\end{align}

The second correction of the same order in $X$
comes from the magnetic dipole interaction, which is enhanced by
the factor of $(g_p-g_n)^2$,
\begin{align}
\delta H =-\frac{e}{2\,m_p}\,\biggl[\sum_a g_a\,\vec s_a-g_d\,\vec s_d\biggr]\,\vec B = -(\vec \mu-\langle\vec \mu\rangle)\,\vec B.\label{26}
\end{align}
The corresponding contribution to the scattering tensor is
\begin{align}
\delta \alpha^{kl} =&
\biggl\langle\bigl[(\vec\mu-\langle\vec\mu \rangle)\times\vec k\bigr]^k
\frac{1}{E_N-\tilde H_N-\omega}\,
\bigl[(\vec\mu-\langle\vec\mu \rangle)\times\vec k\bigr]^l
\nonumber \\ & + (k\leftrightarrow l, \omega\rightarrow -\omega)\biggr\rangle,\label{24}
\end{align}
which gives the following correction to the hyperfine splitting in $\mu$D
\begin{align}\label{eq:Epol2}
  \delta E_{\text{pol2}} =&\  -\frac{\alpha}{16}\frac{m_r^2}{m_p\, m}\, \frac{(g_p-g_n)^2}{g_d}\,E_F
  \Braket{(\vec s_p -\vec s_n)\,X\,(\vec s_p -\vec s_n)}.
\end{align}

Further corrections are of higher order in $X$. The $X^2$ terms vanish, as shown in Ref.
\cite{pachucki_nuclear_2007}, and the next-order nonvanishing terms are proportional to $X^3$.
Namely, the next-order (in $X$) term of the correction coming from $\delta\alpha^{kl}$ in
Eq.~\eqref{eq:vpEpol1} is
\begin{align}
\delta E_{\text{pol3}} = \frac{\alpha}{4}\frac{g_p-g_n}{g_d}E_F
\frac{5m_r-2m }{3\,m^3}m_r^4\Braket{\vec{R}\,X^3\,\vec{R}}.\label{eq:Epol3}
\end{align}

Another $X^3$ correction, $\delta E_{\text{pol4}}$, comes from the following spin dependent
coupling to the electric field \cite{pachucki_nuclear_2007},
\begin{equation}
\delta H = -\vec T \cdot \frac{\partial \vec{E}}{\partial t}\,,
\end{equation}
where $ \vec{T} $ is defined as
\begin{equation}
\vec{T} = \sum_{a}\,\left(\frac{e_a}{2m_p}(g_a-1)-\frac{Z\,e}{2M}\right)\vec{s}_a\times\vec{x}_a\,.
\end{equation}
The corresponding vector polarizability correction is
\begin{align}
  \delta \alpha^{kl} =&\ i\,\omega^3\,
\biggl\langle D^k\,\frac{1}{E_N-\tilde H_N-\omega}\,T^l
+T^l\,\frac{1}{E_N-\tilde H_N+\omega}\,D^k\nonumber \\ &\
-(k\leftrightarrow l, \omega\rightarrow -\omega)\biggr\rangle,\label{36}
\end{align}
and the contribution to the hyperfine splitting of $\mu$D is
\begin{equation}
\delta E_{\rm pol4} = -\frac{\alpha}{3}\frac{g_p-g_n-1}{g_d} E_F \frac{m_r^4}{m^2} \Braket{\vec{R}\,X^3\,\vec{R}}.\label{eq:Epol4}
\end{equation}

The last nuclear polarizability correction $\delta E_{\text{pol5}}$ comes from the fourth-order
terms of the expansion of the vector polarizability $\delta\alpha^{kl}$ in the small parameter
$\vec k\cdot\vec R$. In order to derive this correction, we rewrite Eq.~\eqref{22} by using
$k^i\,T^{ik} = \omega\,T^{0k}$ and apply the nonrelativistic approximation. The result is
\begin{widetext}
\begin{align}\label{eq999}
\delta E_{\rm hfs} =&\ -e^2\,\psi^2(0)\,\int\frac{d\,\omega}{2\,\pi}\,
\int\frac{d^3k}{(2\,\pi)^3}
\frac{(\vec k\times\vec\sigma)^i\,(T^{0i}-T^{i0})}{k^2\,(2\,m\,\omega-k^2)(2\,m\,\omega+k^2)}\nonumber
\\ =&\ \frac{e^4}{m}\,\psi^2(0)\,\int\frac{d^3k}{(2\,\pi)^3}i\frac{(\vec\sigma\times\vec k)^i}{k^4}\,
\sum_{a,b}\biggl\langle
j^0_a\,e^{i\,\vec k\cdot \vec R_a}\,\frac{1}{H_N-E_N+\frac{k^2}{2\,m_r}}\,j^i_b\,e^{-i\,\vec k\cdot\vec R_b}
\biggr\rangle\,,
\end{align}
where, in the nonrelativistic approximation, $j^0_a=1$ and $j^i_b = -i\,\vec\sigma_b\times\vec
k\,g_b/(4\,m_p)$. By neglecting $H_N-E_N$ in the above expression one obtains the Low correction
given by Eq.~(\ref{approx}). The quadratic terms of the expansion of Eq.~(\ref{eq999}) in $\vec
k\cdot\vec R_a$ gives $\delta E_{\rm pol1}$. We now consider the fourth power of  $\vec k\cdot\vec
R_a$ and obtain for $\mu$D the following correction
\begin{align}
  \delta E_{\text{pol5}} =&\ \frac{\alpha}{15}\,E_F\,m_r^4\,\biggl\{ \frac{5}{6}\frac{g_p+g_n}{g_d}\,\Braket{R^2\,X^3\,R^2}
  -2\,\frac{g_p-g_n}{g_d}\,\Braket{R^2\,\vec{R}\,X^3\,\vec{R}}
  \nonumber \\&\
  +\frac{g_p+g_n}{g_d}\,
  \Braket{\left(R^iR^j-R^2\,\delta^{ij}/{3}\right)\,X^3\,\left(R^iR^j-R^2\,\delta^{ij}/{3}\right)}\biggr\}\,.\label{eq:Epol5}
\end{align}
\end{widetext}
We are not aware of any further significant contributions, therefore
we write the nuclear vector polarizability correction as
\begin{equation}\label{eq:Epol}
  \delta E_{\rm pol} = \sum_{i=1}^5 \delta E_{{\rm pol}\,i}\,,
\end{equation}
and assume a $5\%$ uncertainty due to omitted $\delta E_{\rm pol0}$ and higher-order (in $X$)
corrections. Our result disagrees with the result by Faustov and Martynenko
\cite{faustov_hyperfine_2014}, because they used incorrect formula for the polarizability
correction derived for electronic atoms and included in addition the Low correction.

The nuclear vector polarizability is presently the main source of the theoretical uncertainty of
the total nuclear-structure correction. This means that in the future, detailed investigations should
reanalyze all possible contributions to the nuclear vector polarizability.

\section{Three-photon-exchange elastic correction}
The $ (Z\alpha)^2\,E_F$ elastic contribution to the hyperfine splitting $\delta^{(2)}
E_{\text{fns}}$ can be derived by following the approach developed earlier for the case of the Lamb
shift in Ref. \cite{pachucki_three-photon-exchange_2018}. Instead of a direct use of the Dirac
equation, which is possible but tedious, we shall split the total correction into low-energy $
\delta E_L $ and high-energy $ \delta E_H $ parts. Both these parts are separately divergent, so we
employ dimensional regularization with $d=3-2\epsilon$ (see Appendix \ref{app:dimreg} for details)
and cancel singularities $\sim 1/\epsilon$ in the sum $\delta^{(2)} E_{\text{fns}} = \delta E_L +
\delta E_H$. For convenience we assume lepton mass $ m=1 $ from now on and restore it only in the final expression from dimensional analysis.

\subsection{Low-energy part}
In the low-energy part, where $ p\sim Z \alpha $, the nonrelativistic approximation is valid. The
nonrelativistic Hamiltonian $ H $ with the nuclear electric $G_E$ and magnetic $G_M$ formfactors (with their respective Fourier transforms $ \rho_E,\rho_M $),
here normalized to unity, is given by
\begin{equation}\label{eq:H}
H = \frac{p^2}{2}+V(r) +\frac{4\pi Z\alpha }{d m_p}\,g_p(\vec{s}_p\cdot \vec{s}_\mu)_\epsilon\, \rho_M(r)\,,
\end{equation}
where the potential $V$ is defined by its Fourier transform
\begin{equation}
  V(p) = -G_E(p^2)\,\frac{4\pi Z \alpha}{p^2} \label{eq:pot_V}
\end{equation}
and where $ (\vec{s}_p\cdot \vec{s}_\mu)_\epsilon $ is defined in \eqref{reg:spineps}.

Because the characteristic momentum $p$ is much smaller than the inverse of the nuclear size, the
nuclear formfactors can be expanded in $p^2$. We thus obtain
\begin{equation}
  H = H_0 + H_{\rm hfs} + \delta V + \delta H_{\rm hfs}\,,
\end{equation}
where
\begin{align}
  H_0 =&\ \frac{p^2}{2}-Z\alpha \left[\frac{1}{r}\right]_\epsilon, \label{eq:H0} \\
  H_{\rm hfs} =&\ \frac{4\pi Z\alpha }{d m_p}\,g_p(\vec{s}_p\cdot \vec{s}_\mu)_\epsilon\delta^{(d)}(\vec{r}),\label{eq:vm1}\\
  \delta V =&\ - 4\pi Z \alpha \,G_E'(0)\,\delta^{(d)}(\vec{r}), \\
  \delta H_{\rm hfs} =&\  -\frac{4\pi Z\alpha }{d m_p}\,g_p(\vec{s}_p\cdot \vec{s}_\mu)_\epsilon\, G_M'(0)\nabla^2\delta^{(d)}(\vec{r}).
\end{align}
and where $ [1/r]_\epsilon $ denotes a $d$-dimensional generalization of the $ 1/r $
potential. The corresponding correction to the hyperfine splitting of order $ (Z\alpha)^6 $ is
\begin{equation}
  \delta E_L = 2 \Braket{\delta V\,\frac{1}{(E_0-H_0)'}\,H_{\rm hfs}} +  \Braket{\delta H_{\rm hfs}}.
\end{equation}
Calculating matrix elements in $d$ dimensions and using $ G_E'(0) = -r_p^2/6 $, $ G_M'(0) =
-r_{m}^2/6 $, with $ r_p $ and $ r_{m} $ being charge and magnetic radius of the nucleus,
respectively, we obtain the following expression for the low-energy part,
\begin{align}
    \delta E_L =&\frac{4}{d}( r_p Z \alpha)^2E_F\bigg[-\frac{1}{4\epsilon}-\frac{1}{n}-\frac{1}{2}+\gamma-\ln \frac{n}{2}+\nonumber\\
    &+\Psi (n)+\ln (Z \alpha) +\frac{r_{m}^2}{4 r_p^2 n^2} \bigg]\,. \label{eq:El_tot}
\end{align}
The singularity $\sim 1/\epsilon$
in the above equation will cancel out with $ \delta E_H $.

\subsection{High-energy part}
\begin{figure*}[t]
  \includegraphics[width=\textwidth]{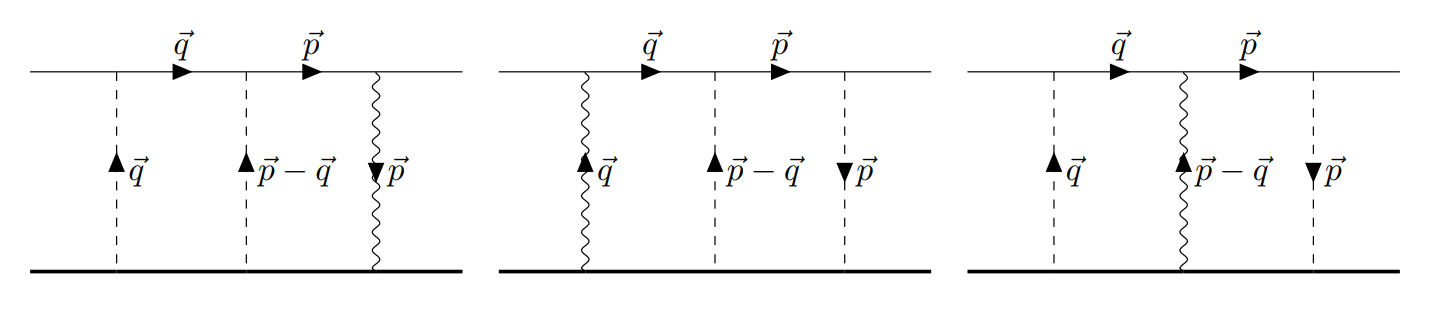}  
	\caption{Three diagrams representing contributions to the high-energy part $ \delta E_H $.
          Wavy lines represent magnetic photons carrying potential \eqref{eq:pot_A},
          and dashed lines are electric photons carrying potential \eqref{eq:pot_V}.}
		\label{fig:diagrams}
\end{figure*}

In the high-energy part $\delta E_H$, the lepton momentum is of the order of the inverse of the
nuclear size, so one can employ the scattering approximation. Specifically, $\delta E_H$ is given
by the forward three-photon exchange amplitude, which can be represented by the three diagrams shown in
Fig.~\ref{fig:diagrams}. The resulting expression is
\begin{align}
\delta E_H =& \delta E_{H1} + \delta E_{H2} +  \delta E_{H3}\label{eq:Eh_1} \\
\delta E_{H1} =&\psi^2(0)\int\frac{d^dp}{(2\,\pi)^d}\,\int \frac{d^dq}{(2\,\pi)^d}\bra{\bar{t}} e\vec{\gamma}\cdot\vec{A}(-\vec{p}) \nn\\
&\times\frac{1}{\slashed{p}-1}\gamma^0V(\vec{p}-\vec{q})\frac{1}{\slashed{q}-1}\gamma^0 V(\vec{q})\ket{t},\nonumber\\
\delta E_{H2} =& \psi^2(0)\int\frac{d^dp}{(2\,\pi)^d}\,\int \frac{d^dq}{(2\,\pi)^d}\bra{\bar{t}}\gamma^0V(-\vec{p})\nn\\
&\times \frac{1}{\slashed{p}-1}\gamma^0 V(\vec{p}-\vec{q})\frac{1}{\slashed{q}-1} e\vec{\gamma}\cdot\vec{A}(\vec{q}) \ket{t},\nonumber\\
\delta E_{H3}=&\psi^2(0)\int\frac{d^dp}{(2\,\pi)^d}\,\int \frac{d^dq}{(2\,\pi)^d}\bra{\bar{t}}\gamma^0V(-\vec{p})\nn\\
&\times\frac{1}{\slashed{p}-1} e\vec{\gamma}\cdot\vec{A}(\vec{p}-\vec{q}) \frac{1}{\slashed{q}-1}\gamma^0 V(\vec{q})\ket{t},\nonumber
\end{align}
where $ t=(1,\vec{0}) $, $ \slashed{p}=(1,\vec{p})^\nu \gamma_\nu $, and
\begin{equation}
	A^i(q) = \frac{i\,Z\,e\,g_p}{4\,m_{p}} \sigma_p^{ik}\frac{q^k}{q^2}G_M(q^2)\,. \label{eq:pot_A}
\end{equation}
After performing Dirac algebra, $ \delta E_H $ can be expressed in the coordinate
representation as
\begin{eqnarray}\label{eq:Eh_2}
\delta E_H &=& \frac{8 \psi^2(0)}{d m_p}(Z\alpha)^3g_p\braket{\vec{s}_p\cdot\vec{s}_\mu}_\epsilon \nonumber\\
&&\times\int d^dr\left(2\pi\rho_M(r)[\mathcal{V}_E^{(2)}]^2 + \mathcal{V}_M\mathcal{V}_E\mathcal{V}_E^{(2)}\right)\,,
\end{eqnarray}
where $ d $-dimensional potentials $ \mathcal{V}_{E} $ and $ \mathcal{V}_{M} $ are defined in
Appendix \ref{app:dimreg}. The first term under the integral sign is convergent due to the presence
of $\rho_M $ and thus can be evaluated in three dimensions. The second term, however, contains a
singularity $\sim 1/\epsilon$, which has to be separated out. This is achieved by splitting the
domain of integration into $ (0,\Lambda) $ and $ (\Lambda,\infty) $. The integral over $
(\Lambda,\infty) $  is evaluated by using the asymptotic form of the potentials in $d$ dimensions.
The final result for the high-energy part is
\begin{equation}\label{eq:Eh_tot}
\delta E_H =  \frac{4}{d} (r_p Z\alpha)^2 E_F\left[ \frac{1}{4\epsilon}+ \ln r_{pp} + \frac{1}{2}
+\gamma \right]\,.
\end{equation}
Here, $ r_{pp} $ is the effective radius defined by
\begin{align}\label{eq:rcc}
  \ln\frac{r_{pp}}{r_p} =& \frac{1}{G_E'(0)}  \int_{0}^{\infty} dr \ln\frac{r}{r_p} \frac{d}{dr} r^3 \bigg\{ 2\pi \rho_M(r) [V_E^{(2)}]^2\nonumber \\
  &+ V_MV_EV_E^{(2)} + \frac{1}{r^2}\left(\frac{r}{2}-\frac{G_E'(0)}{r}\right)\bigg\},
\end{align}
where   $ V_{E(M)} $ are three-dimensional versions of potentials $ \mathcal{V}_{E(M)} $, which depend
on electric and magnetic formfactors, and are presented in Appendix \ref{app:dimreg}.

The final result for the elastic three-photon exchange correction,
$\delta^{(2)} E_{\text{fns}} = \delta E_L + \delta E_H$, with $ m $ restored from dimensional analysis, is
\begin{align}
  \delta^{(2)} E_{\text{fns}} =& \frac{4}{3}E_F(m r_p Z \alpha)^2\bigg[-\frac{1}{n}+2\gamma
    -\ln\frac{n}{2}+\Psi(n)\nonumber \\ &+\ln( mr_{pp}Z \alpha) + \frac{r_{m}^2}{4r_p^2 n^2} \bigg]\,.\label{eq:E_tot}
\end{align}
This expression is valid for any nucleus, both for muonic and electronic atoms.

For the dipole parametrization of the nuclear formfactors
 \begin{equation}\label{eq:formfactor}
 G_E(q) = G_M(q) = \frac{1}{(1+q^2/\Lambda^2)^2}\,,
 \end{equation}
 one easily obtains the following results
 \begin{align}
 	r_{m} =r_p\,,\ \
 	r_{Z}/r_p = \frac{35}{16 \sqrt{3}}\,,\ \
 	r_{pp}/r_p = 5.274\,565\ldots\,,
 \end{align}
which are independent of parameter $ \Lambda $ and thus valid for any nucleus. Numerical values of
the elastic $(Z\alpha)^2\,E_F$ correction for the $ 1S $ and $ 2S $ states of muonic hydrogen are
\begin{align}\label{33}
  \delta^{(2)} E_{\text{fns}}(\mu{\rm H}) = \left\{\begin{array}{ll}
\SI{-0.0093}{\milli\electronvolt}\,,& \mbox{\rm for the $1S$ state}\,,\\
\SI{-0.00096}{\milli\electronvolt}\,,& \mbox{\rm for the $2S$ state}\,.
\end{array}\right.
\end{align}
The above result for the $2S$ state  deviates from the value of $-0.00065$~meV calculated by
Indelicato \cite{indelicato_nonperturbative_2013} and quoted by Antognini
\emph{et~al.}~\cite{antognini_theory_2013} (entry h21 in Table III of that work). One of possible
reasons could be the inclusion of the reduced mass in Indelicato's calculation, while our result is
obtained in the nonrecoil limit. We claim that the recoil effect on this relativistic correction
cannot be accounted for in terms of the reduced mass. Our analytic result is verified by a
numerical calculation in the following subsection.

\subsection{Numerical verification}

The analytical expression \eqref{eq:E_tot} for the fns correction of order $ (Z\alpha)^2 E_F $ has
been verified by comparison with the numerical evaluation of the fns correction to all orders in
$Z\alpha$. Specifically, the numerical all-order fns correction was obtained by evaluating the
expectation value of the Fermi-Breit operator $H_{\mu}$ with solutions of the Dirac equation with
an extended nucleus and subtracting the point-nucleus result. The (extended-nucleus) Fermi-Breit
operator is
\begin{equation}
H_{\mu} = \frac{|e|}{4\pi}\frac{\bm{\alpha}\cdot[\bm{\mu}\times \bm{r}]}{r^3}\,F(r)\,,
\end{equation}
where $\bm{\mu}$ is the nuclear magnetic moment operator, $\bm{\alpha}$ is the vector of Dirac
matrices, and $F(r)$ describes the radial distribution of the magnetic moment, $F(r) \approx 1$
outside of the nucleus. For the dipole parametrization (\ref{eq:formfactor}), the distribution
function is given by
\begin{equation}
F(r) = 1 - e^{-\lambda r}\Big(1 + \lambda r + \frac12 (\lambda r)^2 \Big) \,,
\end{equation}
where $\lambda = 2\sqrt{3}/r_m$.

\begin{figure}[h]
 	\includegraphics[width=\linewidth]{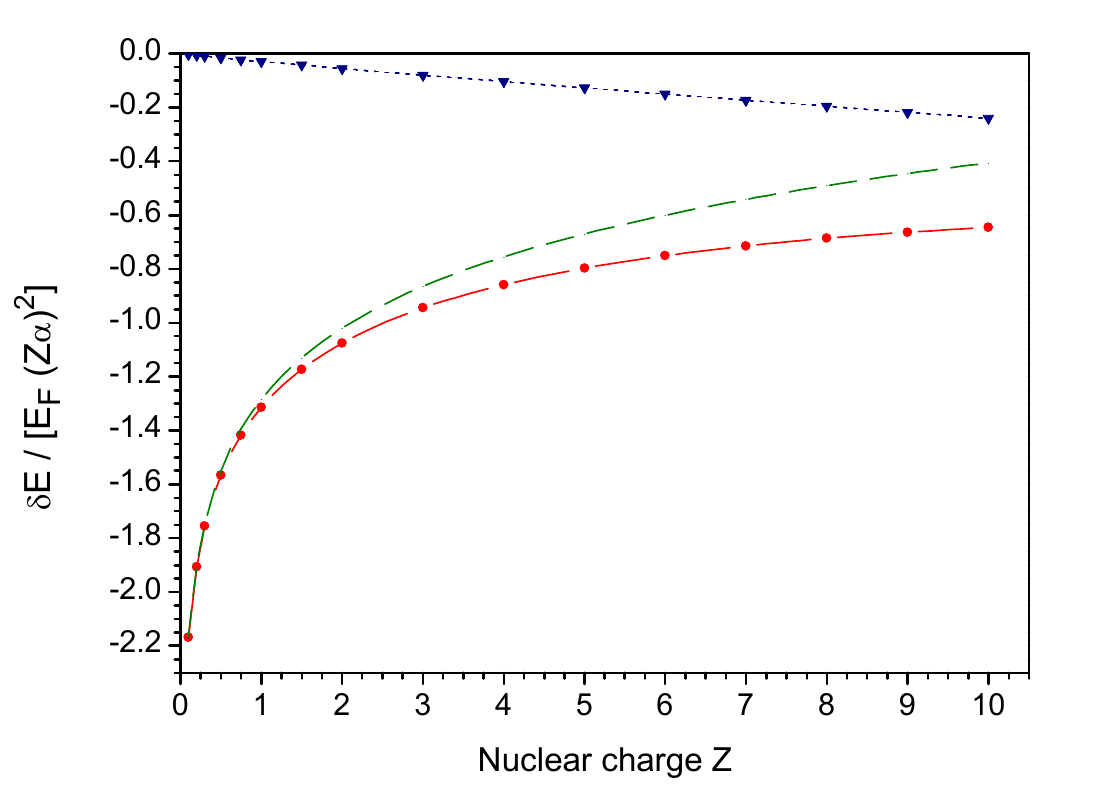}
	\caption{Numerical and analytical results for the fns correction normalized by the factor of $E_F (Z\alpha)^2$,
for the $1S$ state of muonic hydrogen-like ions with the nuclear root-mean-square radius 1~fm, as a function of nuclear charge $Z$. Solid line
and dots (red) represent numerical results (\ref{eq:Eres}); dashed line (green) shows analytical results (\ref{eq:E_tot});
dotted line and triangles (blue) show the difference.}
	\label{fig:num}
\end{figure}

We define the relativistic fns correction that contains orders $ (Z\alpha)^2 E_F $ and higher by
subtracting the Zemach contribution from the numerical fns correction,
\begin{equation}\label{eq:Eres}
\delta^{(2+)} E_{\text{fns}} =  E_{\text{fns,num}} - \delta E_{\text{Zem}}\,.
\end{equation}
Results for the relativistic fns correction are presented in Fig.~\ref{fig:num}, which demonstrates
agreement between the numerical and analytical approaches for $Z=0$. The difference between the
numerical and analytical results scales linearly with $Z$, representing the
contribution of orders $ (Z\alpha)^3 E_F $ and higher.

\section{Three-photon exchange nuclear structure correction}

In this section we address the {\em inelastic three-photon exchange} nuclear-structure correction
$\sim\!(Z\,\alpha)^2\,E_F$ to the hyperfine splitting,
which has not been studied in the literature so far. We will perform an
approximate treatment of this correction and evaluate the largest contribution, namely that due to
the electric dipole polarizability. Furthermore, we will demonstrate partial cancellations
occurring between the elastic and inelastic parts for the case of muonic deuterium, $\mu$D.

The nuclear-structure correction $ \delta^{(2)} E_{\rm nucl} $ can be represented as a sum of several contributions,
\begin{equation}\label{eq:POL_E_def}
\delta^{(2)} E_{\rm nucl} = \delta E_{\text{DP}} + \delta E_L(D) +   \delta E_H(p) + \delta E_H(pn)
\end{equation}
where $ \delta E_{\text{DP}} $ is the low-energy dipole polarizability contribution, $ \delta
E_L(D) $ is the elastic low-energy part, and $ \delta E_H(p) $  and $ \delta E_H(pn) $ are the
high-energy corrections.

It is convenient to introduce the common factor
\begin{equation}\label{eq:POL_Edbar}
\mathcal{E} = \frac{4\pi\,\alpha}{3m_p}\,\psi^2(0)\,,
\end{equation}
which will frequently appear in formulas below.

\subsection{Elastic low-energy contribution}
The elastic low-energy part $ \delta E_L(D) $ can be obtained from Eq.~\eqref{eq:El_tot} by
replacing the proton charge $ r_p $ and magnetic $ r_{m} $ radii with their deuteron counterparts $ r_d $ and $r_{md} $,
because at low energies muon sees the nucleon as a whole. The resulting expression is

\begin{align}
\delta E_L(D) =&\frac{4}{d}( r_d \alpha)^2\mathcal{E}\bigg[-\frac{1}{4\epsilon}-\frac{1}{n}-\frac{1}{2}+\gamma-\ln \frac{n}{2}+
\nonumber\\&
\Psi (n)+\ln\alpha +\frac{r_{md}^2}{4 r_d^2 n^2} \bigg]\,g_d\,\Braket{\vec{s}_d\cdot \vec{s}_\mu}_\epsilon, \label{eq:POL_El_tot}
\end{align}
where $\Braket{\vec{s}_d\cdot \vec{s}_\mu}_\epsilon$ is defined in Eq. (\ref{reg:spineps}).

\subsection{Polarizability contribution}

We derive here the leading dipole polarizability correction $\delta E_{\text{DP}}$
and represent it as a sum of two terms
\begin{equation}
\delta E_{\text{DP}} = \delta E_{\text{DP1}} + \delta E_{\text{DP2}}
\end{equation}
The first one $\delta E_{\text{DP1}}$ is obtained by taking the
second-order matrix element with dipole interaction in the nonrelativistic approximation and
perturbing it with the magnetic dipole-dipole interaction $ H_{\text{hfs}} $. The result is
\begin{align}\label{eq:POL_E_pol_1}
\delta E_{\text{DP1}} =&\ \alpha^2 \delta_{\text{hfs}} \biggl\langle\psi \phi _N\bigg| \vec{R}\cdot\vec\nabla\left[\frac{1}{r}\right]_{\epsilon}
\\ &\ \times \nonumber
\frac{1}{E_N + E_0 - H_N -H_0}\vec{R}\cdot\vec\nabla\left[\frac{1}{r}\right]_\epsilon\bigg|\psi\phi _N\biggr\rangle\,.
\end{align}
Here, $\vec{R}$ is the proton position with respect to the deuteron mass center and
$\delta_{\text{hfs}}$ denotes the first-order perturbation due to the hyperfine interaction $ H_{\text{hfs}} $
which is an analog of $H_{\rm hfs}$ in Eq. (\ref{eq:vm1}) for the deuterium
\begin{equation}
 H_{\text{hfs}} =  \frac{4\pi\alpha}{d m_p}\delta^{(d)}(\vec{r})\, g_d( \vec{s}_d \cdot \vec{s}_\mu)_\epsilon ,\label{eq:POL_H_hfs}
\end{equation}
where $ H_N $ is the nuclear Hamiltonian and $ H_0 $ is the nonrelativistic muon Hamiltonian defined in
Eq.~(\ref{eq:H0}). The perturbative treatment of $H_{\text{hfs}}$ means that the polarizability
correction is expressed as a sum of two terms, originating from perturbations of the denominator
and the wave function. However, the first term vanishes and $\delta E_{\rm DP1}$ becomes
\begin{widetext}
\begin{eqnarray}\label{eq:POL_Eden}
  \delta E_{\text{DP1}} &=& 2\alpha^2 \bigg\langle \psi \phi_N \bigg\rvert
 \vec{R}\cdot\vec\nabla\left[\frac{1}{r}\right]_\epsilon \frac{1}{E_0+E_{N}-H_0 - H_{N}} \vec{R}\cdot\vec\nabla\left[\frac{1}{r}\right]_\epsilon
\frac{1}{(E_0-H_0)'}H_{\text{hfs}}\bigg\lvert \psi \phi_N \bigg\rangle\,,\label{eq:POL_Ewf}
\end{eqnarray}
Because we are interested in the leading correction only, we neglect the $D$-wave in
the ground deuteron state and neglect the Coulomb corrections, so
\begin{eqnarray}
  \delta E_{\text{DP1}} &=&  \frac{12}{d^2}\alpha^2  \mathcal{E} \sum_{\Lambda}
            \bra{\phi_N}\vec{R}\ket{\Lambda}\bra{\Lambda}\vec{R}\ket{\phi_N}
g_d\braket{\vec{s}_d\cdot\vec{s}_\mu}_\epsilon
\int \frac{d^dp}{(2\pi)^d} \frac{d^dq}{(2\pi)^d} \frac{4\pi}{p^2}\frac{4\pi}{q^2}\frac{\vec{p}\cdot\vec{q}}{(\vec{p}-\vec{q})^2}\frac{2}{p^2+2\Lambda}.\label{eq:POL_Ewf2}
\end{eqnarray}
\end{widetext}
After integration we obtain the following expression for the polarizability correction
\begin{equation}
  \delta E_{\text{DP1}} = \frac{4}{d} ( r_s \alpha)^2\mathcal{E}\bigg(\frac{1}{4\epsilon}+\frac{2}{3}-
  \frac{1}{2}\ln 2\bar{E}\bigg)\Braket{g_d\vec{s}_d\cdot\vec{s}_\mu}_\epsilon\\
	    \label{eq:POL_E_pol_tot}\,,
\end{equation}
where $ \bar{E} $ is the mean excitation energy  defined by
\begin{equation}\label{eq:POL_Ebar}
\ln \frac{\bar{E}}{m} =\frac{1}{r_s^2}\Braket{\phi_N|\vec{R}\ln\biggl[\frac{(H_N-E_N)}{m}\biggr]\vec{R}|\phi_N}\,.
\end{equation}
and $ r_s $ is the deuteron structure radius
\begin{equation}\label{eq:POL_rs}
r_s = \sqrt{\Braket{R^2}}\, .
\end{equation}

The mean excitation energy \eqref{eq:POL_Ebar} was calculated in
Ref.~\cite{pachucki_three-photon-exchange_2018} using the AV18 potential \cite{wiringa:95},
with the result
\begin{equation}\label{eq:POL_Ebarval}
\bar{E} = \SI{7.37(7)}{\mega\electronvolt}.
\end{equation}
The pion-less EFT in the next-to-leading order \cite{friar_nuclear_2013} reproduces this result with $ \SI{2}{\percent} $
accuracy.

The second dipole polarizability contribution $ \delta E_{\rm DP2}$ is the Coulomb distortion
correction to the leading two-photon exchange contribution $ \delta E_{\rm pol1} $ in Eq.~\eqref{eq:Epol1}.
We derive it by considering the nonrelativistic formula for the second order correction to energy
that comes from the electric dipole interaction and the magnetic quadrupole term in Eq.~\eqref{eq:dHmag}
\begin{widetext}
\begin{align}
  \delta E_{\rm DP2} =&\  4\pi\alpha^2\,\frac{g_p-g_n}{d m_p}\Braket{\vec{s}_d\cdot\vec{s}_\mu}_\epsilon\delta_C
  \biggl\langle\psi\phi_N|\vec{R}\cdot\vec{\nabla}\left[\frac{1}{r}\right]_\epsilon
    \frac{1}{H_0+H_N-E_0-E_N}\vec{R}\cdot\vec{\nabla}\, \delta(\vec{r})|\psi\phi_N\biggr\rangle,
\end{align}
where $\delta_C$ denotes the Coulomb correction, namely that beyond $E_{\rm pol1}$ in Eq. (\ref{eq:Epol1}).
This Coulomb correction is the forward scattering three-photon exchange amplitude,
which takes the following form in the momentum representation
\begin{align}
  \delta E_{\rm DP2} &= \frac{3 \alpha^2}{d}\frac{g_p-g_n}{g_d}E_F\sum_{\Lambda'}\braket{\phi_N|R^k|\Lambda}\braket{\Lambda|R^i|\phi_N}
  \int\frac{d^dp}{(2\pi)^d}\frac{d^dq}{(2\pi)^d}
  \frac{4\,\pi\,p^k}{p^2}\frac{4\,\pi\,q^i}{\frac{q^2}{2}+\Lambda}
  \biggl[\frac{1}{\frac{p^2}{2}+\Lambda} \frac{1}{\abs{\vec{p}-\vec{q}}^2} + \frac{2}{\abs{\vec{p}-\vec{q}\,}^4}\biggr].
\end{align}
\end{widetext}
After integration we obtain for $ \delta E_{\rm DP2} $
\begin{equation}
\delta E_{\rm DP2} = \frac{4}{3}(\alpha\, r_s)^2\frac{g_p-g_n}{g_d} E_F\left(\frac{3}{4}- \ln 2\right)
\end{equation}
Its value is heavily suppressed by the numerical factor in the parentheses.

\subsection{High-energy contribution $ \delta E_H(pn)$}
When muon momentum is of the order of the inverse of internucleon distance, the muon sees different
positions of the proton and the neutron inside the nucleus, and effectively one can discern which
photon interacts with which nucleon in the three-photon exchange. Because one can neglect the nuclear
excitation energy in comparison to the muon kinetic energy, the high-energy contribution can be
represented as an expectation value of the effective interaction potential,
\begin{equation}
  \delta E_H(pn) = \langle\phi_N|\delta V_H|\phi_N\rangle\,,
\end{equation}
where
\begin{align}
 \delta V_H &= \psi^2(0)\,\int\frac{d^dp}{(2\,\pi)^d}\,\int \frac{d^dq}{(2\,\pi)^d}\bigl[\delta V_{H1} + \delta V_{H2} + \delta V_{H2}\bigr]\,,\label{eq:POL_Eh_1}
\end{align}
and
\begin{widetext}
  \begin{align}
	\delta V_{H1} &=\bra{\bar{t}} e\vec{\gamma}\cdot\vec{A}_a(-\vec{p}) e^{i \vec{p}\cdot\vec{R}_a}\frac{1}{\slashed{p}-1}\gamma^0 V_b(\vec{p}-\vec{q})e^{i (\vec{q}-\vec{p})\cdot\vec{R}_b}\frac{1}{\slashed{q}-1}\gamma^0 e^{-i \vec{q}\cdot\vec{R}_c}V_c(\vec{q})\ket{t},\nonumber\\
	\delta V_{H2} &=\bra{\bar{t}}\gamma^0V_c(-\vec{p})e^{i \vec{p}\cdot\vec{R}_c}\frac{1}{\slashed{p}-1}\gamma^0 V_b(\vec{p}-\vec{q})e^{i (\vec{q}-\vec{p})\cdot\vec{R}_b}\frac{1}{\slashed{q}-1} e^{-i \vec{q}\cdot\vec{R}_a} e\vec{\gamma}\cdot\vec{A}_a(\vec{q}) \ket{t},\nonumber\\
	\delta V_{H3} &=\bra{\bar{t}}\gamma^0V_b(-\vec{p})e^{i \vec{p}\cdot\vec{R}_b}\frac{1}{\slashed{p}-1} e\vec{\gamma}\cdot\vec{A}_a(\vec{p}-\vec{q}) e^{i (\vec{q}-\vec{p})\cdot\vec{R}_a}\frac{1}{\slashed{q}-1} \gamma^0 e^{-i \vec{q}\cdot\vec{R}_c}V_c(\vec{q})\ket{t}.\nonumber
	\end{align}
\end{widetext}

Indices $(a,b,c) \in \{p,n\}$ discern whether the interacting nucleon is the proton or the neutron.
We neglect all the contributions where the electric photon interacts with the neutron ($ b=c=p $) and
thus are left with two cases. We first consider the case when the magnetic photon hits the neutron
($ a=n $). This gives the following correction
\begin{align}
  &\delta E_H(pn) = \frac{8  \psi^2(0)}{dm_p}\alpha^3g_n\braket{\vec{s}_n\cdot\vec{s}_\mu}_\epsilon\label{eq:POL_Eh_2} \\
  & \times \bigg\langle\phi_N\bigg\rvert\int d^dr\bigg(2\pi\rho_{M}(\vec{r}+2\,\vec{R})\,[\mathcal{V}_E^{(2)}]^2\nonumber\\
  &+\mathcal{V}_M(\vec{r}+2\,\vec{R})\,\mathcal{V}_E\mathcal{V}_E^{(2)} \bigg)\bigg\lvert\phi_N\bigg\rangle\nonumber\,,
\end{align}
which is almost the same as Eq. \eqref{eq:Eh_2}, but has magnetic terms shifted by the proton-neutron distance.
All the potentials in Eq. \eqref{eq:POL_Eh_2} are functions of $ r $, unless explicitly written and magnetic potentials refer to the neutron formfactor,
while electric ones to the proton one, both through definitions \eqref{reg:int1}. The treatment of Eq.
\eqref{eq:POL_Eh_2} follows the same pattern as that of Eq.  \eqref{eq:Eh_2}. The only difference
is that magnetic factors are shifted by $2\,\vec R$, which leads to  additional terms. The result is
\begin{align}\label{eq:POL_Eh_3}
\delta E_H(pn) &= \frac{4}{d}( r_p \alpha)^2\mathcal{E}g_n \Braket{ \vec{s}_n\cdot\vec{s}_\mu}_\epsilon\nn\\
&\times\bigg(\frac{1}{4\epsilon}+\gamma-\frac{1}{2}+  \frac{3r_{pn}^2+2r_s^2}{r_p^2}+\ln 2 r_l\bigg),
\end{align}
where $ r_l $ is defined as
\begin{equation}\label{eq:POL_Rbar}
\ln \left(2 m r_l\right) = \braket{\phi_N|\ln \left(2 m R\right)|\phi_N}
\end{equation}
and its value can be found in Table \ref{tab:eff_radii}.
The effective proton-neutron radius $ r_{pn} $ in Eq.~\eqref{eq:POL_Eh_3} is defined as
\begin{align}
  &2\pi{r_{pn}^2}= \bigg\langle\phi_N\bigg\rvert\int d^3r \Bigg(2\pi\rho_{M}(\vec{r}+2\,\vec{R})[V_E^{(2)}]^2 \label{eq:POL_rpn} \\
  &+V_M(\vec{r}+2\,\vec{R})\,V_E\,V_E^{(2)} 
  +\frac{1}{\lvert\vec{r}+2\,\vec{R}\rvert}\,
  \bigg(\frac{1}{2}-\frac{G_E'(0)}{r^2}\bigg)\Bigg)\bigg\lvert\phi_N\bigg\rangle.\nonumber
\end{align}
where $\rho_{M}$ and $V_M$ correspond to  the neutron, while $G_E$ and $V_E$ correspond to the proton.
        It is worth noting that in the point-nucleon limit the proton-neutron effective radius $ r_{pn} $, given by Eq.
        \eqref{eq:POL_rpn},
        is equal to the deuteron structure radius ($ r_{pn}\rightarrow r_s $).

\subsection{High-energy contribution $ \delta E_H(p) $}
In the case when all three photons interact with the proton, one should use the complete
four-vector current, as in the two-photon case. We are not able to perform such a calculation at
present, and thus we assume that the dominating contribution comes from the elastic part in the
nonrecoil limit,
\begin{align}
\delta E_H(p) =&  \frac{4}{d} ( r_p \alpha)^2 \mathcal{E}\left[ \frac{1}{4\epsilon}+ \ln r_{pp} + \frac{1}{2}
  +\gamma \right]\! g_p\Braket{ \vec{s}_p\cdot \vec{s}_\mu}_\epsilon\,.
\end{align}

\subsection{Total three-photon exchange correction}
Summing all contributions in Eq.~\eqref{eq:POL_E_def} and restoring $m$'s from dimensional
analysis, we obtain the three-photon nuclear-structure correction of order $(Z\,\alpha)^2\,E_F$ for
$\mu$D,
\begin{align} \label{eq:POL_fin}
  \delta^{(2)} E_{\rm nucl} =&
  \frac{4}{3} (m \alpha)^2 E_F\bigg[r_{pn}^2\frac{3g_n}{2g_d} +r_{md}^2\frac{1}{4n^2} +r_s^2\bigg(\frac{g_n}{g_d}-\gamma\nn\\
  &+\frac{g_p-g_n}{g_d}\left(\frac{3}{4}- \ln 2\right)+\frac{1}{6}
	-\frac{1}{2}\ln\frac{ 2\bar{E}}{m}\bigg)\nn\\
	& + r_d^2\left(	2\gamma-\frac{1}{n}-\ln \frac{n}{2}+\Psi (n)  +\ln \alpha\right) \nn \\
	& +r_p^2\bigg(  g_p\ln (m r_{pp})+g_n\ln (2 m r_l)-g_n \bigg)\frac{1}{2g_d} \bigg]\,.\nn \\
\end{align}
In deriving this final formula we made use of an approximate identity $\braket{g_d
\vec{s}_d}\approx \braket{g_p \vec{s}_p+g_n \vec{s}_n}/$ to cancel out the
$\epsilon$ singularity and $ \braket{\vec{s}_p}=\braket{\vec{s}_n}=\braket{\vec{s}_d}/2 $ for
further simplifications. Numerical values for all parameters in the final formula are listed in
Table \ref{tab:eff_radii}.

\begin{table}[!htb]
	\caption{Effective radii and $g$-factors for the
three-photon nuclear-structure correction  for $\mu$D, given by Eq.~\eqref{eq:POL_fin}.
Proton-neutron radius $ r_{pn} $ is taken in the point-nucleon limit ($ r_{pn} = r_s $)
and $r_l$ is obtained with the AV18 potential \cite{wiringa:95}}
	\label{tab:eff_radii}
	\begin{ruledtabular}\as{1.25}
		\begin{tabular}{cw{2.6}crw{2.5}}
			\centt{Variable}&   \centt{Value} & \centt{Units} & \centt{Source} \\
			\hline
			$ r_{p} $ &  0.84087(39)&\text{fm}   &  Ref.~\cite{antognini_proton_2013} \\
			$ r_{d} $ & 2.1256(8)&\text{fm} & Ref.~\cite{pohl_deuteron_2017} \\
			$r_{s}$   & 1.954661(79) & \text{fm} & Ref.~\cite{pachucki_three-photon-exchange_2018}\\
			$r_{md} $    & 2.312(10)&\text{fm}   & Ref.~\cite{afanasev_magnetic_1998}  \\
			$r_{pp}$    & 4.435&\text{fm}   &  Eq. \eqref{eq:rcc}     \\
			$r_{pn}$    & 1.955 &\text{fm}   &  Eq. \eqref{eq:POL_rpn}   \\
			$ r_{p\rm Z} $ & 0.883(19) & \text{fm} & Ref.~\cite{toamalak_arxiv_18} \\
			$ r_{n\rm Z} $ & 0.06(1) & \text{fm} & Ref.~\cite{tomalak} \\
			$r_{l}$ & 1.339&\text{fm} & Eq. \eqref{eq:POL_Rbar} \\
			$ g_p $ & 5.585 694 702(17) &   & Ref.~\cite{mohr_codata_2016}\\
			$ g_d $ & 0.8574382311(48) &   & Ref.~\cite{mohr_codata_2016}\\
			$ g_n $ & -3.82608545(90) &   & Ref.~\cite{mohr_codata_2016}\\
		\end{tabular}
	\end{ruledtabular}
\end{table}

Our final results for the three-photon exchange deuteron structure correction $ \delta^{(2)} E_{\rm
nucl} $ are presented in Table~\ref{tab:en_corr}. We assume a \SI{25}{\percent} uncertainty of
these results, due to neglect of the polarizability corrections beyond the electric dipole
contribution and of the unknown inelastic proton contribution, although we must admit that
we can not well justify this uncertainty estimate.

Our results for the three-photon exchange structure correction disagree with corresponding formulas
obtained by Faustov {\em et al.} \cite{faustov_hyperfine_2014} (given by Eqs. (55), (58), and (59)
of that work). The reason for this disagreement is two-fold. First, Faustov {\em et al.} considered
only the low-energy part of the nuclear-structure correction, omitting completely the high-energy
part. Second, the term proportional to the deuteron magnetic radius $ r_{md} $ in their calculation
contains an additional factor of $ 1/n^3 $, due to a mistake in evaluation of the matrix element
presented in Eq. (\ref{A4}).

\section{Summary}
A summary of all known nuclear-structure contributions to the hyperfine splitting of the $1S$ and
$2S$ states in $\mu$D is presented in Table~\ref{tab:en_corr}.
\begin{table}[!htb]
  \caption{Nuclear structure corrections for hyperfine splitting of the $1S$ and $2S$ states of muonic deuterium, in meV.
    Numerical results are obtained with the AV18 potential \cite{wiringa:95}.}
	\label{tab:en_corr}
	\begin{ruledtabular}\as{1.25}
		\begin{tabular}{lw{2.6}w{2.5}l}
			\centt{Correction}&   \cent{1S} & \cent{2S} & \centt{Source} \\
			\hline
                        $ \delta E_{\rm pol1} $    & -1.1007 & -0.1376   & Eq.~\eqref{eq:Epol1}  \\
			$ \delta E_{\rm pol2} $    & -0.0823 & -0.0103   &  Eq.~\eqref{eq:Epol2}     \\
                        $ \delta E_{\rm pol3} $    &   0.1513      &  0.0189  &  Eq.~\eqref{eq:Epol3}     \\
                        $ \delta E_{\rm pol4} $    &  -0.1979   &      -0.0283     &  Eq.~\eqref{eq:Epol4}     \\
                        $ \delta E_{\rm pol5} $    &    -0.0327     &     -0.0041      &  Eq.~\eqref{eq:Epol5}     \\ \hline
			$ \delta E_{\rm pol } $   &  -1.2623(631) & -0.1578(79) &  Eq. \eqref{eq:Epol}   \\
			$ \delta E_{\rm 1nucl} $   & -0.9450(224) & -0.1181(28) & Eq.~\eqref{eq15} \\
                        $ \delta E_{\rm Low} $     & 2.566  &  0.3208  &  Eq.~\eqref{eq:Elow} \\
			\hline
			$ \delta^{(1)} E_{\rm nucl} $  & 0.3587(670) & 0.0448(84) & Eq. \eqref{eq:E2pe} \\
			$ \delta^{(2)} E_{\rm nucl} $ & -0.0547(137) & -0.0065(16) & Eq.~\eqref{eq:POL_fin}\\
			\hline
			$ \delta E_{\rm nucl, theo}$ & 0.304(68) & 0.0383(86)  & Eq.~\eqref{eq:Enucl}\\
                        $ \delta E_{\rm nucl, exp}$ &           &  0.0966(73)  & Eq.~\eqref{eq:Ediff}\\ \hline
                        difference              &           & 0.0583(113)
		\end{tabular}
	\end{ruledtabular}
\end{table}
We find that the total theoretical result for the deuteron structure correction for the $2S$ state
differs from the experimental value by about $5\,\sigma$. A possible reason for this discrepancy
might be our insufficient knowledge of the spin-dependent coupling of nucleus to the
electromagnetic field, in particular the unknown corrections to the nonrelativistic current in Eq. (\ref{eq999}).
Another possible reason could be a mistake in calculations of QED effects
for the point nucleus, although it looks much less probable since these calculations were performed
independently by two groups \cite{borie:14, faustov_hyperfine_2014}.

Summarizing, in the present work we have calculated the two- and three-photon exchange nuclear
structure corrections to the hyperfine splitting of the $nS$ states in $\mu$D. The obtained results
disagree with the previous theoretical calculation \cite{faustov_hyperfine_2014} and with the
experimental result \cite{pohl:16}.

\begin{acknowledgments}
        We wish to thank Oleksandr Tomalak for helpful comments.
	Part of this work has been performed during MK's stay at the Institut f\"{u}r Kernphysik,
	Johannes Gutenberg-Universit\"{a}t Mainz. MK expresses his gratitude to Sonia Bacca
        for the support and hospitality. This work was supported by the National Science Center (Poland)
        Grant No. 2017/27/B/ST2/02459.
        V.A.Y. also acknowledges support from
   the Ministry of Education and Science of the Russian Federation Grant No. 3.5397.2017/6.7.
\end{acknowledgments}

\appendix
\section{Dimensional regularization}\label{app:dimreg}
In order to extend spin $ \frac{1}{2} $ into $ d $ dimensions, we define antisymmetric tensor
\begin{equation}
\sigma^{ij} = \frac{i}{2}[\gamma^i,\gamma^j]\, ,
\end{equation}
that in the three-dimensional limit simplifies to
\begin{equation}
\sigma^{ij}\overset{d\rightarrow 3}{=} 2\varepsilon^{ijk}s^k.
\end{equation}
 Additionally, for the deuteron we define $ \sigma_d^{ij} = \sigma_p^{ij}+\sigma_n^{ij} $
and use the following convenient notation
\begin{align}
	(\vec{s}_a\cdot \vec{s}_b)_\epsilon&=\frac{1}{8}\sigma_a^{ij}\sigma_b^{ij}\label{reg:spineps}
\end{align}
Throughout our calculations we extensively used the following result for the general $ d
$-dimensional integral,
\begin{equation}\label{eq:Vd}
\int \frac{d^dp}{(2\pi)^d} \frac{4\pi}{p^n} e^{i\vec{p}\cdot\vec{r}} = 2^{2-n} \pi^{1-d/2}\frac{\Gamma\left(\frac{d-n}{2}\right)}{\Gamma\left(\frac{n}{2}\right)}r^{n-d}\,.
\end{equation}
Two special cases are of particular importance (with $d = 3-2\epsilon$):
\begin{eqnarray}
	\mathcal{V}(r) &= \int \frac{d^dp}{(2\pi)^d} \frac{4\pi}{p^2} e^{i \vec{p}\cdot\vec{r}} &= \pi^{\epsilon-1/2}\Gamma(1/2-\epsilon) \frac{1}{r^{1-2\epsilon}}, \nonumber\\
	\label{eq:V2_def}
	\mathcal{V}^{(2)}(r) &= \int \frac{d^dp}{(2\pi)^d} \frac{4\pi}{p^4} e^{i \vec{p}\cdot\vec{r}} &= \frac{1}{4}\pi^{\epsilon-1/2}\Gamma(-1/2-\epsilon) r^{1+2\epsilon}.\nonumber
\end{eqnarray}
We define also the associated potentials:
\begin{eqnarray}
	\mathcal{V}_X &=& 4\pi \int \frac{d^dp}{(2\pi)^d} \frac{G_X(p^2)}{p^2} e^{i \vec{p}\cdot\vec{r}}, \nonumber\\
	\mathcal{V}_X^{(2)} &=& 4\pi \int \frac{d^dp}{(2\pi)^d} \frac{G_X(p^2)}{p^4} e^{i \vec{p}\cdot\vec{r}},\label{reg:int1}
\end{eqnarray}
where $ X\in \{E,M\} $. Asymptotic forms of these potentials at $ r\rightarrow\infty $ are:
\begin{eqnarray}
	\mathcal{V}_X &\rightarrow& \mathcal{V} + \text{local terms}, \nonumber\\
	\mathcal{V}_X^{(2)} &\rightarrow& \mathcal{V}^{(2)} +  G_X'(0)\mathcal{V} + \text{local terms}\nonumber\,.
\end{eqnarray}
The corresponding three-dimensional potentials are
\begin{align}
	V_X =&\ 4\pi \int \frac{d^3p}{(2\pi)^3} \frac{G_X(p^2)}{p^2} e^{i \vec{p}\cdot\vec{r}}, \nonumber\\
	V_X^{(2)} =&\ 4\pi \int \frac{d^3p}{(2\pi)^3} \frac{[G_X(p^2)-1]}{p^4} e^{i \vec{p}\cdot\vec{r}},
\end{align}
with the asymptotic form
\begin{eqnarray}
	V_X &\rightarrow& \frac{1}{r} + \text{local terms},\nonumber\\
	V_X^{(2)} &\rightarrow& -\frac{r}{2} + \frac{G_X'(0)}{r}+\text{local terms}.
\end{eqnarray}
Furthermore, we point out that in the framework of dimensional regularization the following matrix
element vanishes
\begin{equation}
\Braket{\vec{p}~\delta^{(d)}(\vec{r})~\vec{p}} = 0.\label{A4}
\end{equation}

\end{document}